\documentclass[aps,twocolumn,superscriptaddress]{revtex4-2}

\usepackage{amsmath,amssymb}
\usepackage{mathrsfs}
\usepackage{graphics,graphicx}
\usepackage{dcolumn,bm}
\usepackage{breqn}
\usepackage{psfrag}
\usepackage{color}
\usepackage{appendix}
\usepackage[normalem]{ulem}
\usepackage{enumitem}

\usepackage{amsmath,amssymb}
\usepackage{graphics,graphicx}
\usepackage{dcolumn,bm}
\usepackage{psfrag}
\usepackage{enumerate}

\newcommand{\cu}
{\affiliation{Department of Physics, University of Calcutta,
92 Acharya Prafulla Chandra Road, Kolkata 700009, India.}}
\newcommand{\iiser}
{\affiliation{Indian Institute of Science Education And Research Kolkata
Mohanpur, Nadia 741 246
West Bengal, India}}

\begin{document}
\title{Virtual walks in the Ising model: finite time scaling}
\author{Amit Pradhan}
\author{Parongama Sen}
\cu
\author{Sagnik Seth}
\iiser

\date{\today}
\begin{abstract}
 The dynamics of the spins  in the  Ising model are analyzed using a virtual walk 
scenario. The system is quenched from a very high temperature
to a lower one using the Glauber scheme in one and two dimensions. A walk is associated with each spin which 
evolves according to the current state of the spin. The probability distribution of the displacement is calculated that shows a distinct change as the temperature is increased. The average displacement as a function of time shows a non-equilibrium region stretched over a much longer time interval compared to the bulk magnetization. Nevertheless, one can still detect a time dependent critical point determined by two different methods. In addition, we introduce a virtual walk constructed from the local energy of individual spins. Finite time scaling of the different quantities estimated in two dimensions show excellent consistency with the values of the known critical exponents.

\end{abstract}
\maketitle

\section{Introduction}

 Phase transitions in the Ising model have been  extensively studied since its first result in  one dimension was published  exactly a century back. 
  Much later, the  two dimensional model was solved exactly and at present, the 
 equilibrium behavior in all dimensions is  well understood.

Most of the recent works on the Ising model are focused on its non-equilibrium behavior. One of the most important phenomena that has been studied 
extensively in the Ising model and its variants is  quenching, where the system is allowed to evolve
from a homogeneous initial state corresponding to a very high temperature to a symmetry broken final state at a temperature less than the critical temperature \cite{Bray1,Privman,Puri,Fierro}. This takes place by the process of  domain coarsening \cite{Suman,Munkel,Saikat,Subir,Menyhard,Henrik,Denis,Nalina,Satya}.

 A zero temperature quench using Glauber kinetics \cite{Glauber} in the Ising model revealed several interesting features. The persistence probability that estimates the
 tendency of spins to remain in their initial state showed the existence of 
 a new exponent in all dimensions \cite{Derrida-PRL,Bray}.
In the two dimensional Ising-Glauber model,  the discovery of the existence of frozen striped phases occurring in a finite number (nearly $30\%$) of cases 
\cite{Spirin,D L Stein} led to  various subsequent studies on different  lattices, initial conditions, boundary conditions \cite{Liu,Tikader,Henkel,Newman} etc. 
At finite temperatures, there are several other features of interest.  
Close to the critical point one encounters critical slowing down;
here the timescale shows a power law divergence where the associated dynamical critical exponent can differ for models with the same equilibrium behavior \cite{Hohenberg}.

In some previous studies \cite{Drouffe,Dornic,Godreche,Baldassarri},  the behavior of the average local spins 
over time was analysed and several other measures of  persistence probability related to this quantity were defined. 
The studies also considered finite temperatures. 

 The sum of the spin at a particular lattice site over time can alternatively be
regarded as a displacement of a walker in a virtual space till that time.
In a virtual space of one dimension,  for  the $i$th spin, the 
displacement $x_i$ of the corresponding walker is given by
\begin{equation}
 x_i(t+1)=x_i(t) + \xi_i (t+1),  
 \label{walk-def}
\end{equation}
where $\xi_i(t)$ is simply equal to $S_i(t)$, the state of the $i$th spin at time $t$.
The probability distribution $P(x,t)$  of the displacement $x$ of the walker at time $t$,  obtained from the walks of all the spins had been estimated which  showed a characteristic change from a two peaked structure to a single peaked Gaussian as the critical point is crossed. Arguing that above the critical point, the states of the spins would be random and the resultant walk will be a random walk, it is easy to explain the Gaussian behavior. The detection of the critical point in this manner is far less complicated than that done using finite size scaling. However, usually there is an overestimation due to the finite
size effect. 

Recently, such virtual walks have been defined in other systems as in econophysics models, generalized voter model in two dimensions and in kinetic exchange   models of opinion dynamics \cite{arnab,goswami2,Pratik,P.sen,Surajit,Surajit1,goswami}.  For each such system, $\xi$ in Eq. \ref{walk-def} is taken  to be the relevant variable in the model. For models which show a continuous phase transition, it was possible to detect the transition point as well as some other features related to the probability distribution of the displacements.

In this work, we explore the virtual walks in the Ising model in both one and two dimensions in more detail. In both cases, the
critical points are exactly known. 
We find a definite advantage of studying the virtual walks.  While finite size scaling is a well known and standard method of detecting critical temperatures and exponents, it involves numerical simulations for several system sizes which are evolved up to considerable time steps. We show here that the virtual walks help in  estimating  critical temperature as well as the static and dynamic  exponents indirectly,  using the method of finite time scaling from a single system size.

 In addition to studying the spin-walk, we also consider a similar walk involving the local energy at time $t$. This leads to an independent estimation of the critical exponents again using finite time scaling.

  The rest of the paper is organized as follows. In section II, we define the model and dynamics, the results are presented  in the next three sections and finally we draw the conclusions in the last section. Detailed calculations are provided in Appendices A-C.

\section{Spin dynamics and simulation details}

The nearest-neighbor Ising Hamiltonian is
\begin{equation}
    H = -J \sum_{\langle ij \rangle} S_iS_j -B\sum_iS_i,
    \label{ising_H}
\end{equation}
where $S_i = \pm 1$ are the Ising spins located at every site of a $d$ dimensional lattice ($d=1$ and $2$ considered in the present work)  and interactions, of strength $J$,  take place between nearest neighbors. In the simulations, we take  the external field $B=0$ and $J=1$.
 The  known exact critical values of the temperature (in units of $\frac{J}{K_B}$) are zero in one dimension and close to 2.269 in two dimensions.

We numerically simulate the Ising model on $L^d$  lattices using periodic boundary conditions. 
 We start with a completely random configuration which corresponds to a very high temperature.
The system is then quenched to different temperatures using  Glauber kinetics.  One Monte Carlo step (MCS) comprises $L^d$ updates. In each update, a spin is chosen randomly and its state is updated. The results are in general reported for $L = 4096$ in one dimension and  $L=128$ in two dimensions. The maximum number of configurations over which averaging had been done was $6000$ in one dimension and $5000$ in two dimensions. 

To each of the spin  in the  lattice sites  we associate a walker.
 As the $i$th spin $S_i$ evolves with time, the  walker associated with that spin is allowed to move in a  one dimensional virtual  space according to Eq \ref{walk-def} with $\xi(t+1) = S_i(t+1)$. 
 Thus the convention is that the movement towards right (left)  occurs when $S_i =1$ ($S_i =-1$).

In the present work, our aim has been to extract as much information of the system as possible from the nature of the probability distribution $P(x,t)$ of displacement $x$ of the walkers  at time $t$ at different temperatures $T$. At $T = 0$ we checked the consistency of our results with the already known results \cite{Bray,Blanchard,Yurke,Sire,Drouffe,Stauffer,P.M.C,Das}. For example, the persistence probability can be identified as $P(x=\pm t,t)$.

The important quantity from which all results are obtained is $P(x,t)$ which is 
expected to follow a general form
\begin{equation}
    P(x,t) \sim t^{-\lambda} f(x/t^{\lambda}),
    \label{scaling_form_of_P(x,t)}
\end{equation}
where $\lambda$ is the exponent relevant for the virtual walk. 
The exponent $\lambda$ is determined by the nature of the walk which depends on the temperature. 
In both dimensions, we first estimate the critical temperature $T_c$ obtained for the system size considered by a so called ratio method. In two dimensions we additionally employ an independent method to determine $T_c$ and then proceed to calculate various quantities as a function of $|T-T_c|$.  

For the two dimensional model, we define a second walk, the so called  energy-walk which is 
generated by calculating
\begin{equation}
y_i(t+1) = y_i(t) + \xi_i^\prime(t+1),
\label{energy-walk}
\end{equation}
where $\xi_i^\prime(t) = E_i(t) =-S_i(t)\sum_{j\in NN}S_j(t) $, a quantity that can be identified as the local interaction energy of the site $i$. The corresponding distribution is written as $Q(y,t)$. This virtual walk is also one dimensional but here
the step lengths can vary between $\pm 4$, taking integer values at any time.
   
\section{ Results for the spin walk in the 1-$d$ Ising model}

\begin{figure}[h]
\includegraphics[width=3. cm,angle = -90,keepaspectratio]{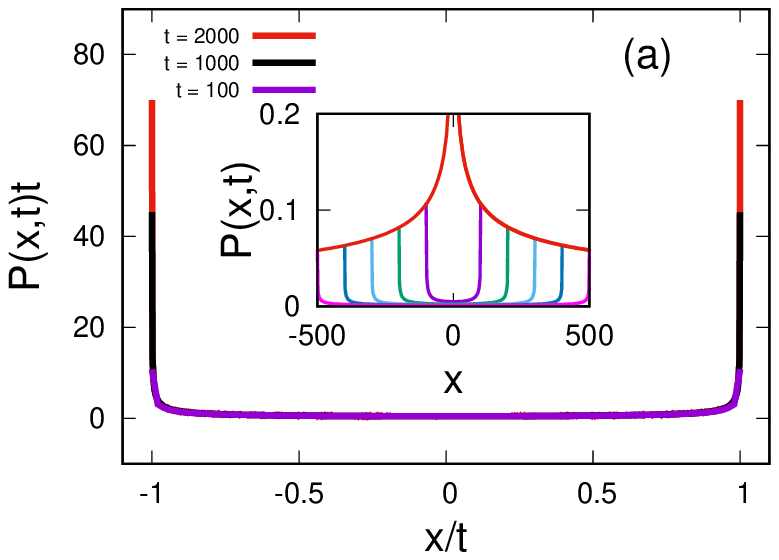} 
\includegraphics[width=3. cm,angle = -90,keepaspectratio]{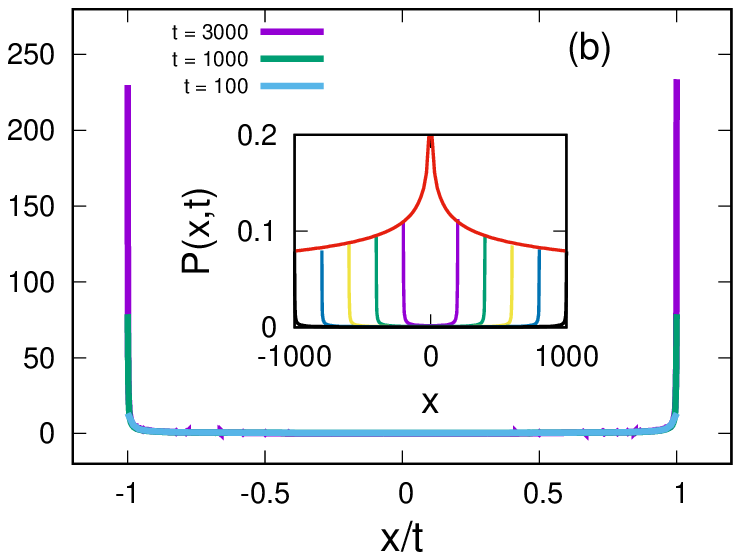} 
\caption{Data collapse of $P(x,t)  $ at three different times shown for $T = 0$ shown for
(a) one dimension and (b) two dimensions. Insets show U-shaped zero temperature probability distribution at different times (t) where the  peaks at  \(x=\pm t\) and $P(x=\pm t,t)$ curves are fitted in the form of $\sim t^{-\theta}$ with $\theta_{1d} = 0.375$ and  $\theta_{2d}\approx 0.2$.}
\label{1d2dT=0}
\end{figure}

Although all results, including persistence probability,  in the one dimensional Ising model are 
known exactly, we explore the walk picture to extract some relevant information. For this, we first check
that the distributions $P(x,t)$ at $T=0$ are U shaped distributions and the data for $P(x,t)t$ at different times collapse when plotted against $x/t$. The scaling of $x \sim t$ indicates that the resultant walk is ballistic in nature. The behavior of $P(x,t)$ at $T=0$  is expected to involve the persistence exponent $\theta$ according to \cite{Drouffe} as,

\begin{equation}
    tP(x,t) \sim (1 - x^{2}/t^{2})^{\theta - 1}.
\label{distT=0}
\end{equation}
 We get a good agreement with the known value of $\theta$ in one dimension; $\theta_{1d} = 0.375$ (see Fig. \ref{1d2dT=0}a). Also, one can directly fit the tips (i.e., the 
points corresponding to $x = \pm t$) of the unscaled distributions at different times in the form $\sim t^{-\theta_{1d}}$ (see inset of Fig. \ref{1d2dT=0}a).

\begin{figure}[h]
\includegraphics[angle = -90, width=9 cm,keepaspectratio]{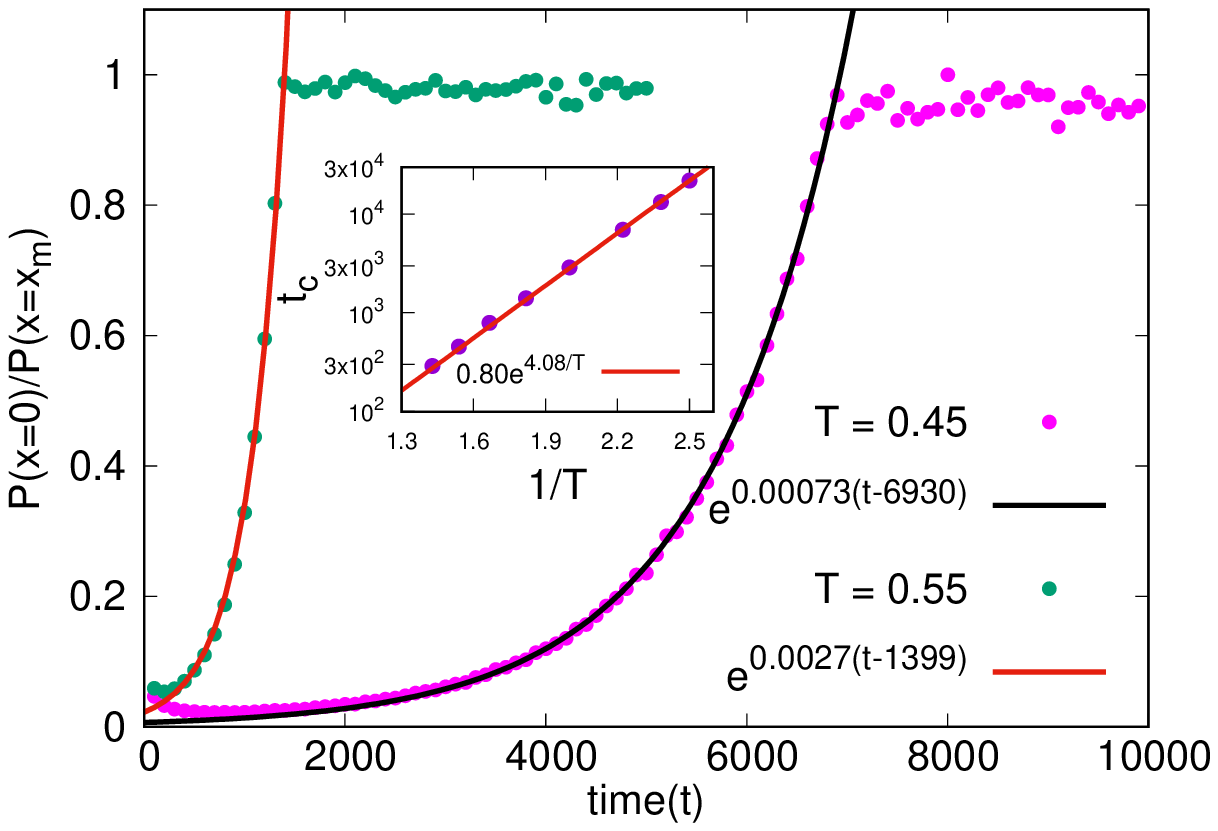} 
\caption{Crossover behavior of the probability distribution $P(x,t)$, from a double-peaked to single-peaked form as time increases, is shown for $T = 0.45,0.55$. The ratio $r = \frac{P(x=0)}{P(x=x_{m})}$ fitted with  exponential form $ e^{a^{\prime}(t-t_c)}$ for $t=t_c^{-}$ with  characteristic time scale $1/a^{\prime} \approx 1370,370  $ and $t_c \approx 6930,1399$ at $T=0.45,0.55$ respectively and $r$ remains $1$ at $t\geq t_c$. Here $t_c$ is the crossover time at which the double peaked structure changes to single peaked structure. Inset shows the variation of crossover time $t_c$ with $\frac{1}{T}$ in log-linear scale. $t_c$ has a exponential divergence at $T=0$ and then it decreases with increasing temperature exponentially fast as $\sim e^{k/T}$ with $k \approx 4.08$.}
\label{Crossover1d}
\end{figure}

At non-zero temperatures, we expect $P(x,t)$ to show a single peaked behavior at $x=0$ as the system becomes disordered. However, at finite times, one still gets a bimodal form.
We analyze the ratio $r = \frac{P(x=0,t)}{P(x=x_m,t)}$ at any temperature where  $x_m$ is the value of $x$ where the probability is maximum (when a double peak exists, we take $x_m>0$,
utilising the symmetry of $P(x,t)$). 
Plotted against time, the ratio goes to the expected value $\approx 1$, only beyond a transient time  $t_c$. Below $t_c$, $r$ is fitted to an exponential form $\exp[a^{\prime}(t-t_c)]$ as we know that at $t_c$, the ratio should be unity. The transient time $t_c$ is proportional to  $\exp{(k/T)}$,
where $k$ is a constant of appropriate dimension, indicating a critical slowing down near the exact critical temperature $T_c = 0$ (shown in the inset of Fig.\ref{Crossover1d}) from which we conclude that the exact critical point can be obtained here from this analysis.

\begin{figure}[h]
\includegraphics[ width=\linewidth]{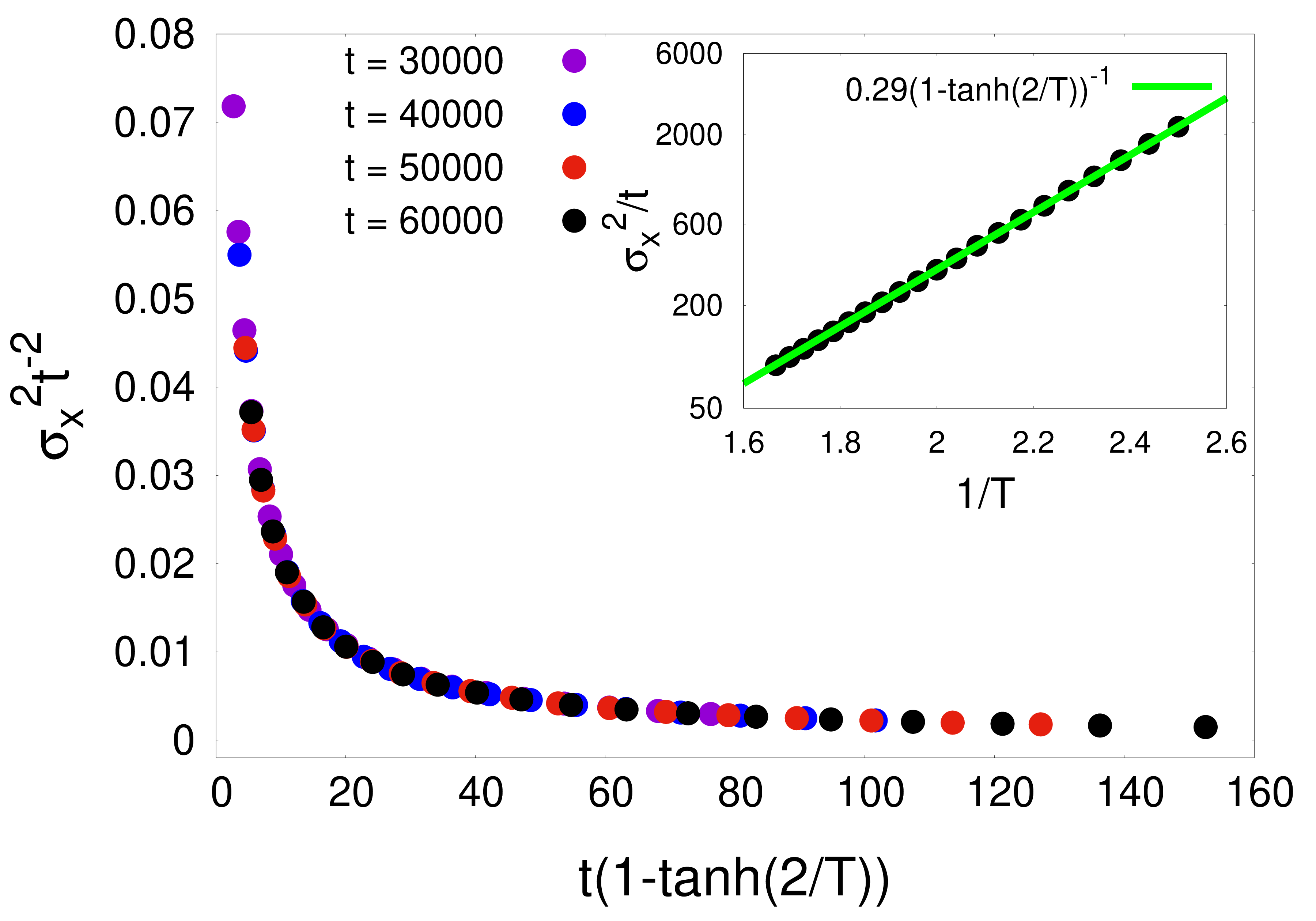} 
\caption{Data collapse of the rescaled variance $\sigma_x^2/t^{2}$ plotted against the scaled variable $t(1-\tanh(2/T))$ for the $1D$ spin walk, confirming the finite time scaling form given in Eq. (\ref{1D FTS}). The data corresponds to fixed observation times $t=30000,40000,50000$ and $60000$. In inset fluctuation per unit time $\sigma_x^2/t$, plotted as a function of inverse temperature $1/T$ in log-linear scale and fitted with the analytical form given in Eq. (\ref{1D spin walk fluctuation}).  }
\label{1D_width_T}
\end{figure}

For the 1-$d$ Ising chain, it is known that instead of power law variations near $T_c=0$, one gets exponential variations for diverging quantities such as characteristic timescales, susceptibility, correlation length etc. As shown above, the characteristic timescale $t_c$ indeed diverges exponentially as $T\to 0$. We now consider the associated one dimensional spin walk fluctuation.   From the cumulative sum in Eq. (\ref{walk-def}), the displacement of the walker associated with a given spin at time $t$ is defined as 
\begin{equation}
\label{Definition of x}
    x_i(t) = \sum_{l=1}^{t} S_i(l).
\end{equation}
The fluctuation of the displacement is therefore given by
\begin{equation}
\label{FLUCTUATION}
    \sigma_x^2(t) = \langle x^2(t)\rangle - \langle x(t)\rangle^2.
\end{equation}
It is possible to estimate the two spin correlations (see Appendix \ref{FTS of fluctuation}) and the known form of the two time spin autocorrelation function for the $1D$ Ising model at $T>0$ \cite{Glauber,Brey}, one finds that $\sigma_x^2(t)$ obeys a finite time scaling form [Eq. (\ref{FTS_1D_variance}) of Appendix \ref{FTS of fluctuation}.1.a]
\begin{equation}
\label{1D FTS}
    \sigma_x^2(t) = t^{2}\mathcal{I}_1\left(\frac{t}{\xi_t}\right)=t^{2}\mathcal{I}_1\left(\alpha t(1-\tanh\left(\frac{2J}{K_BT}\right)\right),
\end{equation}
with scaling function $\mathcal{I}_1$, as derived in Appendix \ref{FTS of fluctuation}.1.a [Eq. (\ref{1D scaling function})].

In the asymptotic regime $t\gg \xi_t$, where $\xi_t$ is the correlation time defined in Eq. (\ref{1D correlation time}) of Appendix \ref{FTS of fluctuation}.1.a, the fluctuation per unit time follows the analytical form [Eq. (\ref{spin_walk_fluctuation_1D}) of Appendix \ref{FTS of fluctuation}.1.a]
\begin{equation}
\label{1D spin walk fluctuation}
   \frac{\sigma_x^2(t)}{t} \sim \xi_t \sim \frac{1}{\left[1-\tanh\left(\frac{2J}{K_BT}\right)\right]}, \quad t \gg \xi_t.  
\end{equation}
Near $T\to 0$, using the expansion $\tanh(x) \approx 1-2\exp(-2x)$ for large $x$, it follows from Eq. (\ref{1D spin walk fluctuation}) that the fluctuation per unit time diverges exponentially as 
\begin{equation}
    \frac{\sigma_x^2(t)}{t} \sim \exp\left(\frac{4J}{K_BT}\right).
\end{equation}
 The finite time scaling relation in Eq. (\ref{1D FTS}) is numerically confirmed by the data collapse shown in Fig. \ref{1D_width_T}.

\section{Results for the spin walk in the $d=2$ Ising model}

 In two dimension, the Ising model manifests a finite temperature phase transition. 
In the next two subsections we present the results for $T=0$ and $T\neq0$ respectively.

To estimate the critical temperature, we analyze the ratio $r = \frac{P(x=0,t)}{P(x=x_m,t)}$ at a particular long time (to avoid any transient effects) at
different temperatures. As already discussed for the one dimensional case, numerically, this ratio should be nearly equal to one above $T_c$ and deviate from 
unity below $T_c$.  Thus one can obtain the value of $T_c$ by fitting $r$  by a suitable function which involves $T_c$  such that $r=1$ at $T_c$. 

 One can also determine the critical temperature by calculating the walk Binder cumulant defined by
\begin{equation}
    U(T,t) = 1 - \frac{\langle x^4 \rangle}{3\langle x^2 \rangle^2}
    \label{binder}
\end{equation}
where the averages are over the displacement distribution $P(x,t)$ at a fixed observation time $t$. Note that usually one defines the Binder cumulant in terms of the bulk order parameter moments.

 In subsections IV.C.1-IV.C.3 and in section V, we consider several quantities and use finite time scaling to extract  both static and dynamic critical exponents .

\subsection{ $T= 0$}

At $T=0$ the results are very similar to those in one dimension. A data collapse can be obtained for $P(x,t)$ curves at different times ($t$) according to Eq.~\ref{scaling_form_of_P(x,t)}   with $\lambda = 1$ [Fig. \ref{1d2dT=0}b]. So, the virtual walk in spin space at $T = 0 $ is a ballistic walk for which the displacement of the walker $x$ scales with time as $ t$ as in the one dimensional case.
The zero temperature probability distribution is extremely close to a beta distribution \cite{Drouffe} and the form involves the persistence exponent $\theta$ as in Eq. \ref{distT=0}.
 Fitting $P(x,t)$ to this form, we obtain the value of the exponent $\theta-1 \approx -0.8$, which gives an estimate of the persistence exponent to be
$\approx 0.2$ which is close to the value found in literature \cite{Bray,Sire,Blanchard,Yurke,Drouffe,Stauffer,P.M.C,Das}. The inset of Fig. \ref{1d2dT=0}b shows the direct fitting of the tips 
to the power law form.

\subsection{ $T> 0$}

 \subsubsection{Determination of $T_c$ using ratio method}

\begin{figure}[h]
\centering
\includegraphics[width=\linewidth]{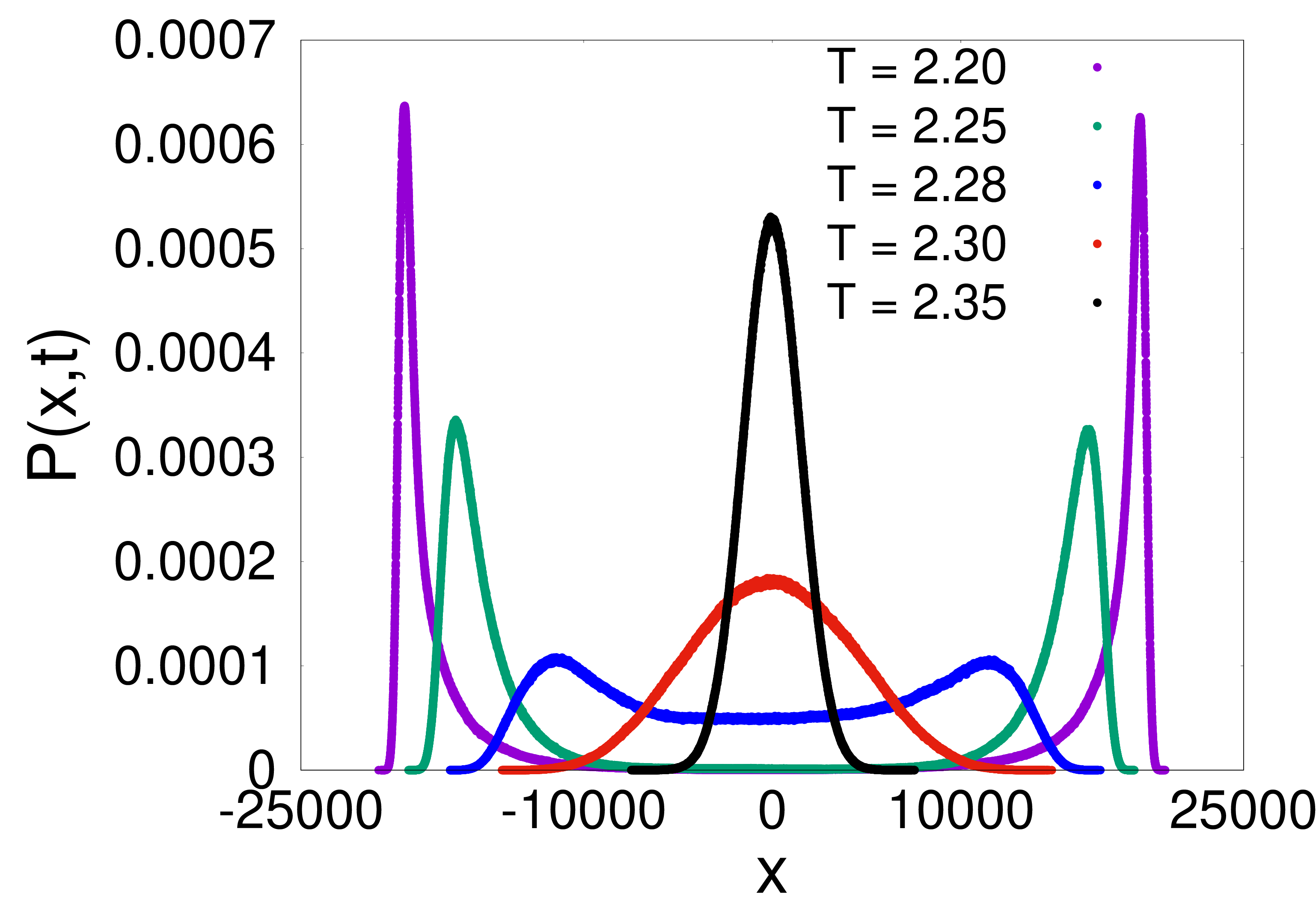} 
\caption{Crossover behavior of the probability distribution $P(x,t)$, from a double-peaked to single-peaked form as temperature increases, is shown for $t = 25000$.}
\label{Crossover}
\end{figure}

Fig. \ref{Crossover} shows the behavior of $P(x,t)$ at different temperatures. 
As $T$ increases from absolute zero, the distribution $P(x,t)$ at long times shows a crossover from a two peaked form to a single peaked form. This should occur as the critical point is crossed. However, from the numerical simulations, a sharp transition from the two peak behavior to the single peak one is difficult to detect, as the peaks become very shallow. So,  we  find the value of $T_c$ indirectly using the ratio
$r=\frac{P(x=0)}{P(x=x_{m})}$.

As $T$ increases, $r$ initially grows with temperature and then saturates to the value $r=1$ at higher temperatures. Arguing that $r=1$ for $T > T_c$, we use a trial function $e^{\mu(T-T_c)}$ to fit $r$ at lower values of $T$, with both $\mu$ and $T_c$ treated as fitting parameters. 

\begin{figure}[h]
\includegraphics[width=\linewidth]{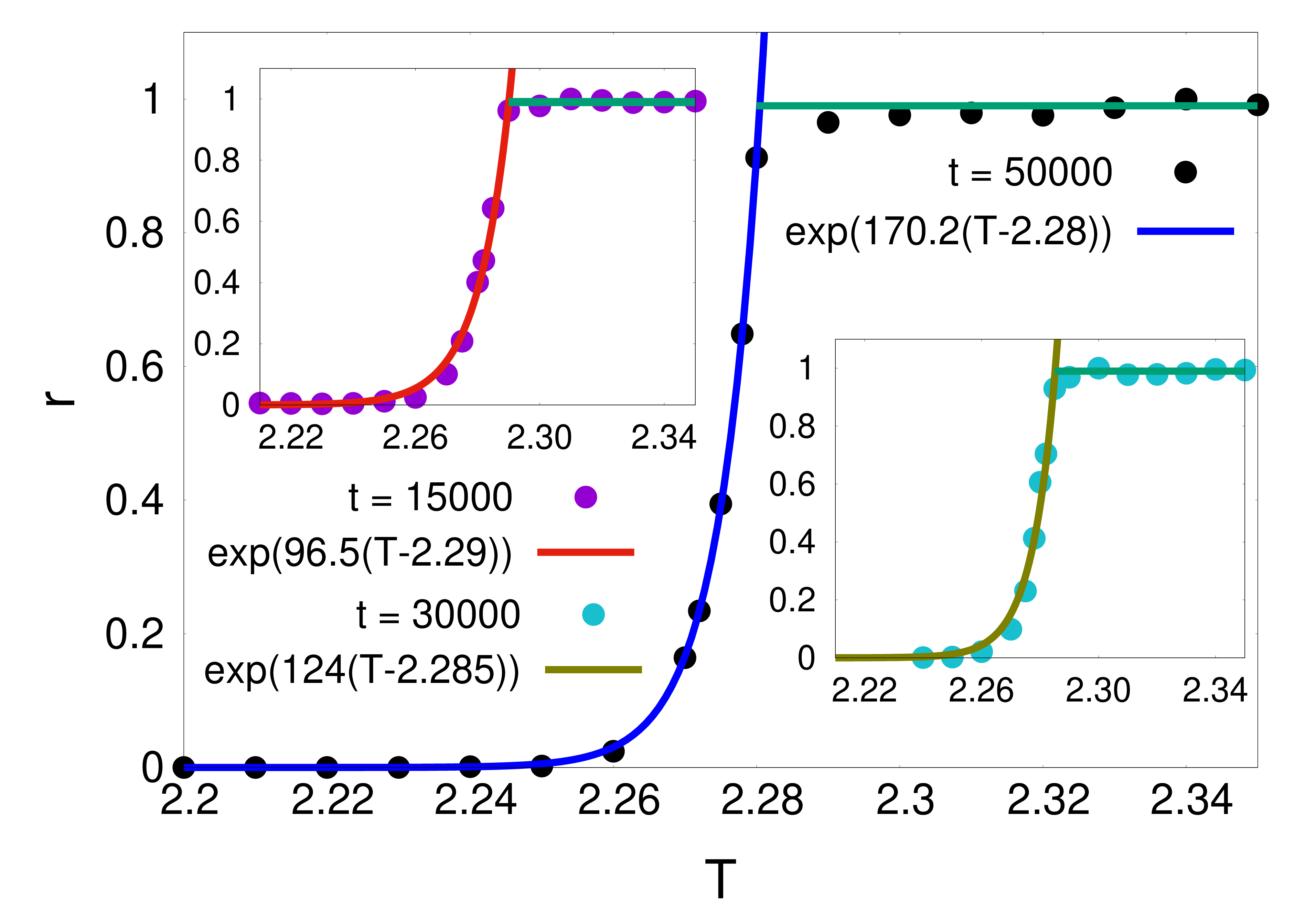} 
\caption{ Plot of  $r = \frac{P(x=0)}{P(x=x_{m})}$  vs $T$ fitted with  exponential form $ e^{\mu(T-T_c)}$ at $T=T_c^{-}$ with  characteristic inverse temperature scale $\mu \approx 96.5,124,170.2  $ and $T_c \approx 2.29,2.285,2.28 $ at $t = 15000,30000,50000$ respectively. $r$ remains $1$ at $T\geq T_c$.}
\label{Determination_of_Tc}
\end{figure}

Fig. \ref{Determination_of_Tc} shows the variation of $r$ as a function of $T$ for several times 
$t$. The fitted parameters $T_c$ and $\mu$ depend on the observation time. The estimated value of $T_c$ decreases with time initially and eventually saturates to $T_c\approx 2.28$ for the largest time considered. We note, however, that this value of $T_c$ for the system size studied represents a slight overestimation of the true critical temperature, which is evidently a finite size effect.

\vskip 0.3cm

 \subsubsection{Determination of $T_c$ from the walk Binder cumulant}
\label{binderc}

\begin{figure}[h]
\includegraphics[ width = \linewidth]{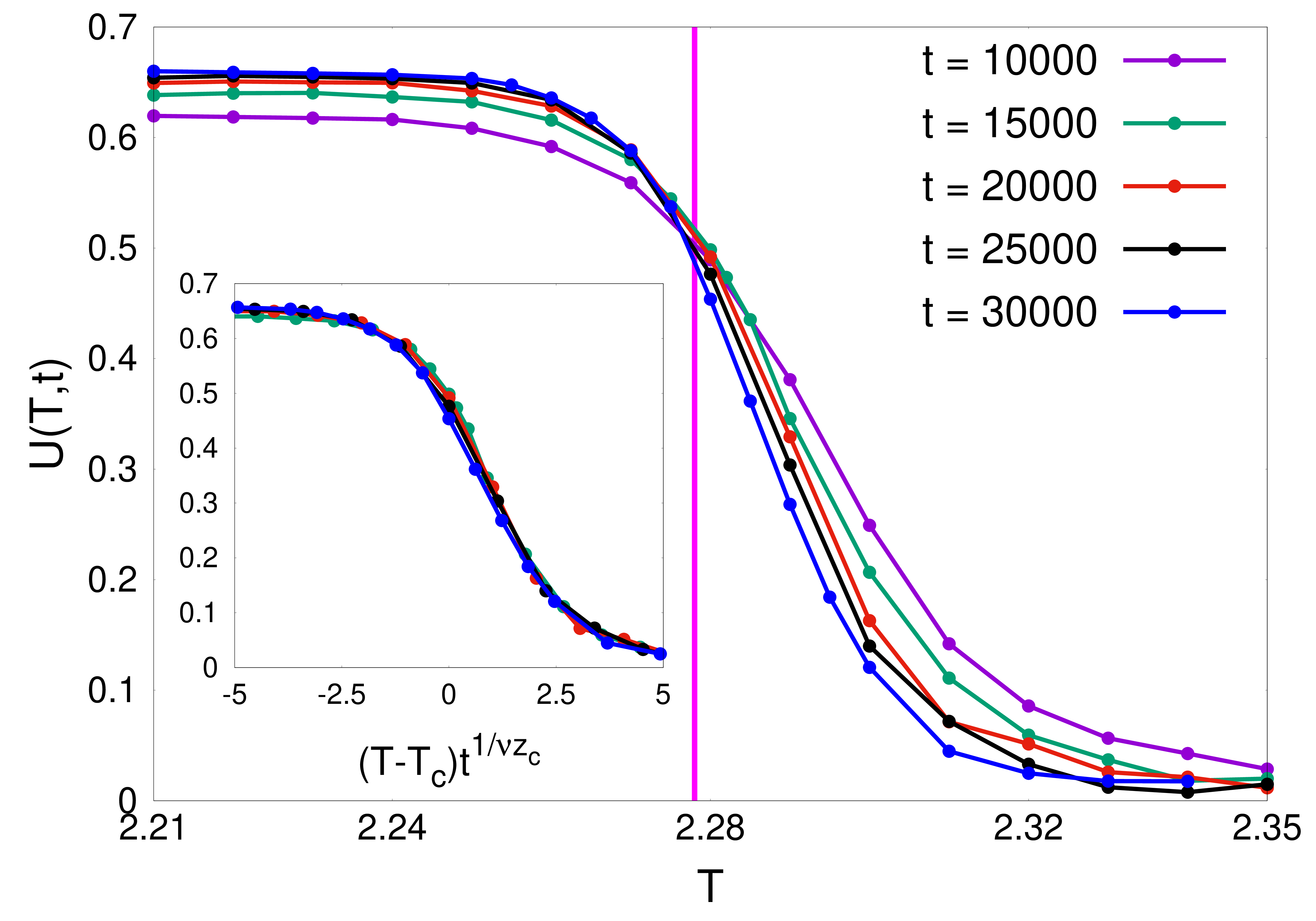} 
\caption{Binder cumulant of the spin walk displacement, $U(T,t)$ is  plotted as a function of $T$ for a system of size $L=128$ at five different evolution times $t=10000,15000,20000,25000$ and $30000$. The curves intersects at a temperature $T\approx 2.28$, providing an estimate of the critical temperature. Inset shows the data collapse of the Binder cumulant using the scaled variable $(T-T_c)t^{1/\nu z_c}$. All curves collapse onto a single universal curve with the expected exponent $1/\nu z_c \approx 0.46$, confirming the finite time scaling form proposed in Eq. (\ref{FTS of binder cumulant}). }
\label{BC vs T}
\end{figure}

An alternative approach To determine $T_c$ is based on the Binder cumulant (Eq. \ref{binder}), constructed from the walker displacements.  When $U(T,t)$ is plotted as a function of temperature for different times $t$, the Binder cumulant curves approach a common crossing point. The crossing point of the curves for different times gives the value of the critical point, $T_c\approx 2.28$ [Fig. \ref{BC vs T}], which is in good agreement with the value of $T_c$ obtained earlier using the ratio method.

\subsubsection{Effective order parameter  $x_m/t$ in the region $T=T_c^-$}
 
In the walk picture,  
in the ordered phase at $T<T_c$, a tagged spin spends most of the time along the same direction and hence the corresponding walk  is ballistic $(\langle x^2\rangle \sim t^2)$ such that $|x_m|$, the most probable displacement is nonzero. In the disordered phase at $T>T_c$, spin  flips are frequent and the walker becomes diffusive $\langle x^2\rangle \sim t$, and $x_m$ vanishes, as shown in Fig. \ref{Crossover}. 
Therefore  $x_m(t)$ distinguishes the ordered and disordered  regimes. 
In the following, we show that $x_m$ will follow same the critical behavior as the usual order parameter, the bulk magnetisation.

 We define the temporal mean of the tagged $i$th spin at time $t$ as
\begin{equation}
    \bar{S_i}(t) = \frac{x_i(t)}{t} = \frac{1}{t}\sum_{\tau=1}^{t}S_i(\tau).
\end{equation}
Averaging this quantity over all sites gives
\begin{equation}
\left\langle \frac{x(t)}{t} \right\rangle_{sites} = \frac{1}{N} \sum_{i=1}^{N} \frac{x_i(t)}{t}=\frac{1}{t} \sum_{\tau=1}^{t} m(\tau),
\end{equation}
where $m(\tau) = \frac{1}{N} \sum_{i=1}^{N} S_i(\tau)$ is the instantaneous bulk magnetization per spin at time $\tau$. Thus, the site averaged temporal mean is nothing but  the time averaged magnetization at all time $t$.

\begin{figure}[h]
\includegraphics[angle = -90, width = 8.5 cm, keepaspectratio]{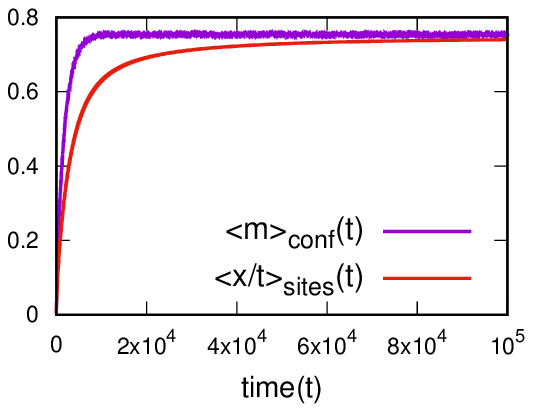} 
\caption{Comparison of the time evolution of the configuration averaged magnetization $\langle m \rangle_{conf}(t) = \frac{1}{N_{conf}} \sum_{i=1}^{N_{conf}}m_i(t)$, where the sum runs over configurations with $m_{eq}(T)>0$ ($m_{eq}(T)$ is the value of magnetization per spin in equilibrium steady state at temperature $T$), and  the site averaged temporal mean $\left\langle \frac{x(t)}{t} \right\rangle_{sites} = \frac{1}{N} \sum_{i=1}^{N} \frac{x_i(t)}{t} $. The close agreement between the two quantities at large times demonstrates that in the steady state the walk order parameter $x_m/t$ coincides with the conventional magnetization order parameter $m_{mp}$. Consequently, the walk order parameter $x_m/t$ should vanish at the critical temperature $T_c$ exactly in a same way as equilibrium magnetization i.e., as $(T_c-T)^{1/8}$.}
\label{avg_m_vs_avg_x}
\end{figure}

The system reaches its steady state at any temperature $T$ after a transient time, say, $t_{eq}$. For $t > t_{eq}$,  the time averaged and the ensemble averaged magnetization coincide. Consequently, the site averaged temporal mean becomes equal to the configuration averaged magnetization [Fig. \ref{avg_m_vs_avg_x}], i.e.
\begin{equation}
 \left\langle \frac{x(t)}{t} \right\rangle_{sites} \xrightarrow[t>t_{eq}(T)]{} \langle m \rangle_{conf}.  
\end{equation}

  Therefore we can consistently define the distributions of $x$ and $M/N = m$ at $t>t_{eq}(T)$ and relate them directly as follows :
 Since the distributions satisfy, $P(x,t)dx = P_m(m,t)dm$ with $x/t = m$ at large time, we obtain
 \begin{equation}
  P(x,t) = \frac{1}{t} P_m\left(\frac{x}{t},t\right).   
 \end{equation}
 Differentiating $P(x,t)$ with respect to $x$ and setting the derivative to zero at the peak gives
 \begin{equation}
     \frac{dP(x,t)}{dx}\big|_{x=x_m} = 0 \implies \frac{dP_m(m,t)}{dm}\big|_{m=\frac{x_m}{t}}=0,
 \end{equation}
so that
\begin{equation}
\label{walk order parameter}
    \frac{x_m(t)}{t} \to m_{mp}(t),
\end{equation}
where $m_{mp}$ is the most probable magnetization per spin. In equilibrium steady state $m_{mp}(t) \to m_{eq}(T)$, the spontaneous magnetization. Therefore in the long time steady state one has $\frac{x_m}{t}(T) \to m_{eq}(T)$, $m_{eq}(T)$ vanishes at $T_c$ as $(T_c-T)^{\beta}$ with equilibrium magnetization exponent $\beta = 1/8$. In the same way, the scaled most probable displacement $x_m/t$ at large time $t$ vanishes at $T_c$ as $(T_c-T)^{\beta}$ with $\beta = \frac{1}{8}$  as shown in the inset of Fig. \ref{Data collapse of xm}.

\vskip 0.3cm
\subsection{Finite time scaling of different quantities}

\subsubsection{Binder cumulant}
  Binder cumulant $U(T,t)$ defined in Eq. (\ref{binder}) is a dimensionless quantity. Using the coarse graining and homogeneity arguments presented in Appendix B, we obtain the finite time scaling form of the Binder cumulant [Eq. (\ref{FTS_binder}) of Appendix B] as
\begin{equation}
\label{FTS of binder cumulant}
    U(T,t) = \mathcal{F}\left[(T-T_c)t^{\frac{1}{\nu z_c}}\right]=\mathcal{F}\left(\frac{t}{\xi_t}\right) ,
   \end{equation}
where  the equilibration time $\xi_t$ diverges as $\xi_t\sim (T-T_c)^{-\nu z_c}$, with $\nu$ the correlation length exponent and $z_c$ the dynamical critical exponent.

 At $T=T_c$, the argument of the scaling function vanishes and $U$ becomes independent of $t$, giving rise to the observed intersection point, $T_c\approx 2.28$.

In simulations, using the scaling variable $(T-T_c)t^{1/\nu z_c}$, with the best known exponent values  $z_c\approx 2.17$ and $\nu=1$ (exact) in two dimensions (i.e., $1/\nu z_c \approx 0.46$), one obtains  a collapse of all the Binder cumulant curves [Inset of Fig. \ref{BC vs T}], confirming the validity of the proposed scaling form
and showing consistency with the known values of the exponents.

 \subsubsection { Walk order parameter  $x_m/t$}

We now focus on the finite time scaling of the walk order parameter $x_m/t$. Close to the critical point, it is expected to obey the finite time scaling form [Eq. (\ref{FTS of x_m/t}) of appendix C]

\begin{equation}
\frac{x_m}{t}(T,t) = t^{-\frac{\beta}{\nu z_c}} \mathcal{G}\left[(T_c-T)t^{\frac{1}{\nu z_c}}\right].
\label{finite time scaling of xm/t}
\end{equation}
where $\beta,\nu$ and $z_c$ are the static and dynamic critical exponents of the Ising universality class. The detailed scaling argument leading to this form is presented in Appendix C.

\begin{figure}[h]
\includegraphics[ width = \linewidth]{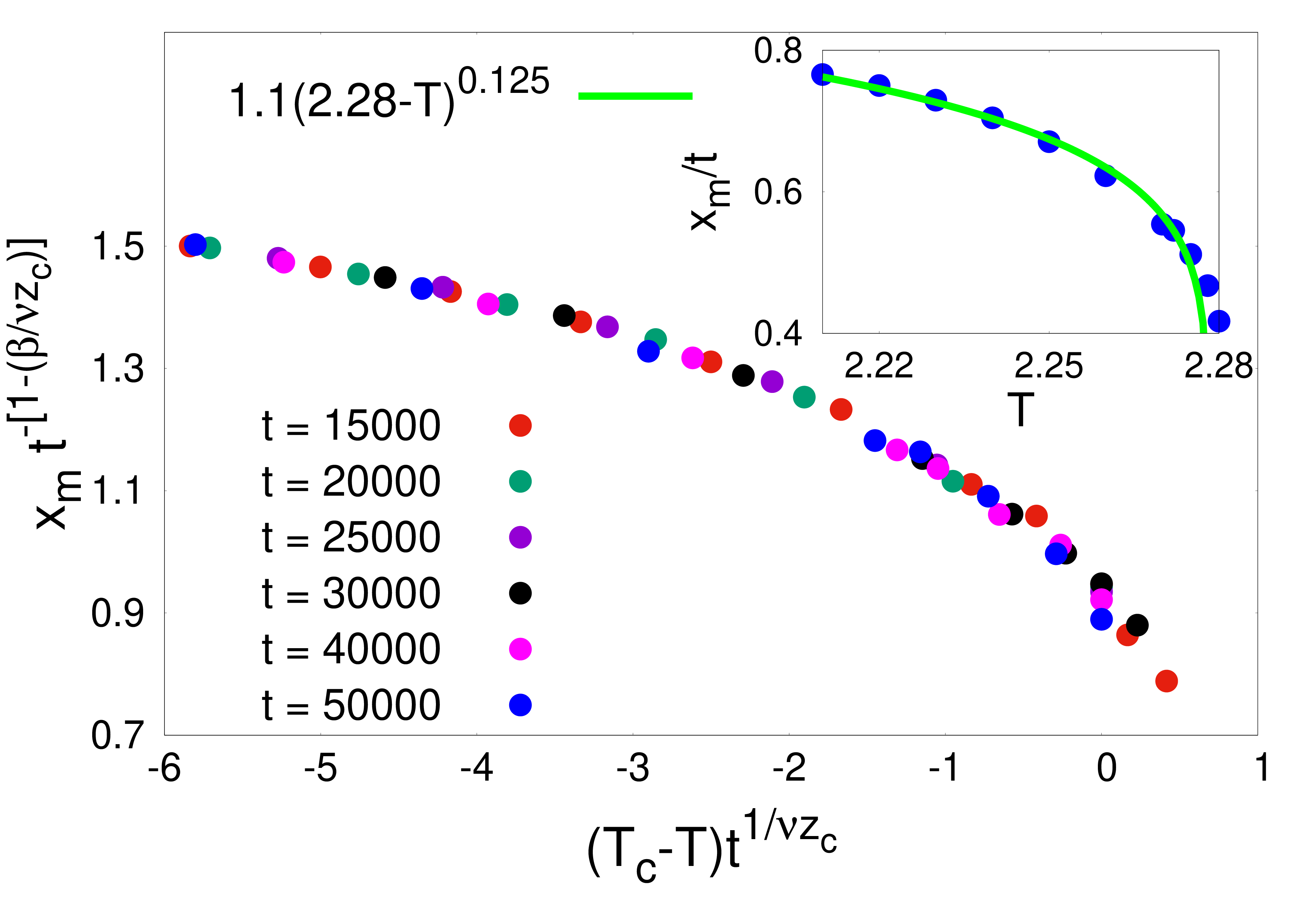} 
\caption{Data collapse of the walk order parameter $x_m/t$ according to the finite time scaling form given in Eq. (\ref{finite time scaling of xm/t}).The rescaled quantity $x_m/t^{1-\beta/\nu z_c}$ with $1-\beta/\nu z_c\approx 0.942$ is plotted against the scaling variable $(T-T_c)t^{1/\nu z_c}$ with $1/\nu z_c\approx 0.46$ for six different observation time $t=15000,20000,25000,30000,40000,50000$. The curves for different $t$ collapses onto a single universal curve, demonstrating that the walk order parameter obeys the predicted finite time scaling given in Eq. (\ref{finite time scaling of xm/t}). Inset shows the most probable value of the local magnetization $x_m/t$ vanishes as $(T_c-T)^{\beta}$ with $\beta =0.125$ at $t=50000$, identical to the behavior of the equilibrium global magnetization. }
\label{Data collapse of xm}
\end{figure}

Using the above variation to apply finite time scaling, we find that indeed the curves at 
different times collapse when the variables are scaled appropriately,  
shown in Fig. \ref{Data collapse of xm}.

\subsubsection{ Diverging fluctuations for spin walk}

 We now focus on the fluctuation of the walker displacement $x$, quantified by the second cumulant $\sigma_x^2(t)$, i.e. the variance of the displacement evaluated over the walk distribution at time $t$. In analogy with the finite size scaling of susceptibility in the Ising system, we analyze the finite time scaling relation of this fluctuation.

 In two dimensions, we use the definition of the spin walk displacement $x_i(t)$ and its fluctuation $\sigma_x^2(t)$ as introduced in Sec. III for the one dimensional case [See Eqs. (\ref{Definition of x}),(\ref{FLUCTUATION})].
Using the definition of $x_i(t)$ in Eq. (\ref{Definition of x}), at large time in the scaling regime assuming stationarity, which is valid only for $T>T_c$ (See Sec. 2.1 of Ref. \cite{Drouffe}) this can be written as 
\begin{equation}
\sigma_x^2(t) \approx 2\sum_{\tau=1}^{t} (t-\tau)C_S(\tau),
\end{equation}
where $C_S(\tau) = \langle S(0)S(\tau)\rangle$, is the two time auto-correlation function of the local spin. The scaling form of $C_S(\tau)$ follows from the general dynamic scaling hypothesis for a local observable (see Eq. (\ref{spin autocorrelation}) of Appendix \ref{FTS of fluctuation}.1.b). Using the scaling dimension of the local Ising spin, $x_S = (d-2+\eta)/2$ (See Chapter 3, Sections 3.7 and 3.8 of Ref.\cite{Cardy}), in an Ising model of spatial dimension $d>1$, one can obtain a finite time scaling form for the spin walk fluctuation at $T\geq T_c$ [Eq. (\ref{FTS of x}) of Appendix \ref{FTS of fluctuation}.1.b.] as
\begin{equation}
\label{finite time scaling of sigma_x^2}
    \sigma_x^2(t,T) \approx t^{\left[2-\left(\frac{d-2+\eta}{z_c}\right)\right]} \mathcal{I}\left[(T-T_c)t^{\frac{1}{\nu z_c}}\right],
\end{equation}
where the scaling function $\mathcal{I}$ arises from an integral over the rescaled time variable $u = \tau/t$, and it is explicitly given in the Appendix \ref{FTS of fluctuation}.1.b [Eq. (\ref{scaling function I})].

 Above the critical temperature, the two time spin autocorrelation becomes exponential (See section 2.1 of Ref.\cite{Drouffe}), As a consequence, the integral defining $\mathcal{I}$ is dominated by contributions from small values of the rescaled time variable  $u$. As a result, as the critical point is approached from $T=T_c^+$, the spin walk fluctuation per unit time at large time [Eq. (\ref{divergence exponent for spin walk}) of Appendix \ref{FTS of fluctuation}.1.b.] diverges as
\begin{equation}
\label{divergence of spin walk fluctuation}
   \frac{\sigma_x^2}{t}\sim (T-T_c)^{-\left[\nu z_c-\nu(d-2+\eta)\right]}.  
\end{equation}
For the two dimensional Ising model, the static and dynamic exponents are $d=2,\nu=1,z_c\approx 2.17,\eta = 1/4$. Using these values, the divergence exponent governing the spin walk fluctuation becomes $\nu z_c-\nu(d-2+\eta) \approx 1.92$.

\begin{figure}[h]
\includegraphics[ width = \linewidth]{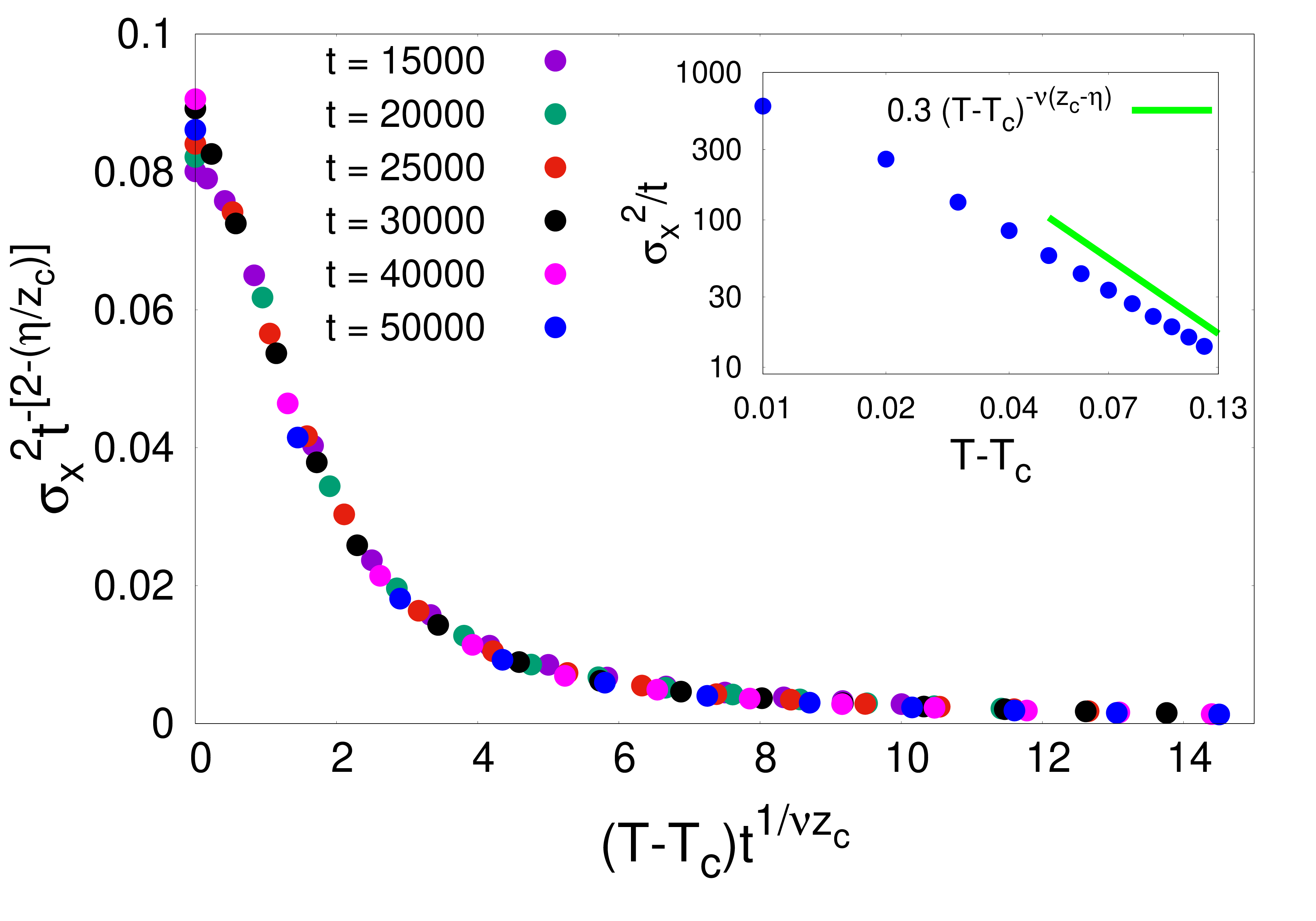} 
\caption{Data collapse of the scaled fluctuation $\sigma_x^2t^{-[2-(\eta/z_c)]}$ plotted against the scaled variable $(T-T_c)t^{1/\nu z_c}$ with $1/\nu z_c \approx 0.46$ and $\eta/z_c\approx 0.115$ at $T>T_c$. Results for six different evolution times $t=15000,20000,25000,30000,40000,50000$ collapse onto a single universal curve, confirming the validity of the finite time scaling relation proposed in Eq. (\ref{finite time scaling of sigma_x^2}). In inset, fluctuation of the spin walk displacement per unit time, $\sigma_x^2/t$, plotted as a function of $T-T_c$ on a  log-log scale at $t=50000$. As the critical point is approached from above, $\sigma_x^2/t$ diverges as $(T-T_c)^{-\nu(z_c-\eta)}$, and the slope extracted from the data is consistent with the expected exponent $\nu(z_c-\eta) \approx 1.92$ [Eq. (\ref{divergence of spin walk fluctuation}]). }
\label{fluctuation of x vs T}
\end{figure}

 The validity of finite time scaling relation given in Eq. (\ref{finite time scaling of sigma_x^2}) is confirmed numerically, as shown in Fig. \ref{fluctuation of x vs T}, where the simulation data collapse onto a single universal curve when plotted according to the proposed scaling form.

\section{Results for the energy walk in the $d=2$ Ising model}

\begin{figure}[h]
\includegraphics[angle = -90, width = 9 cm, keepaspectratio]{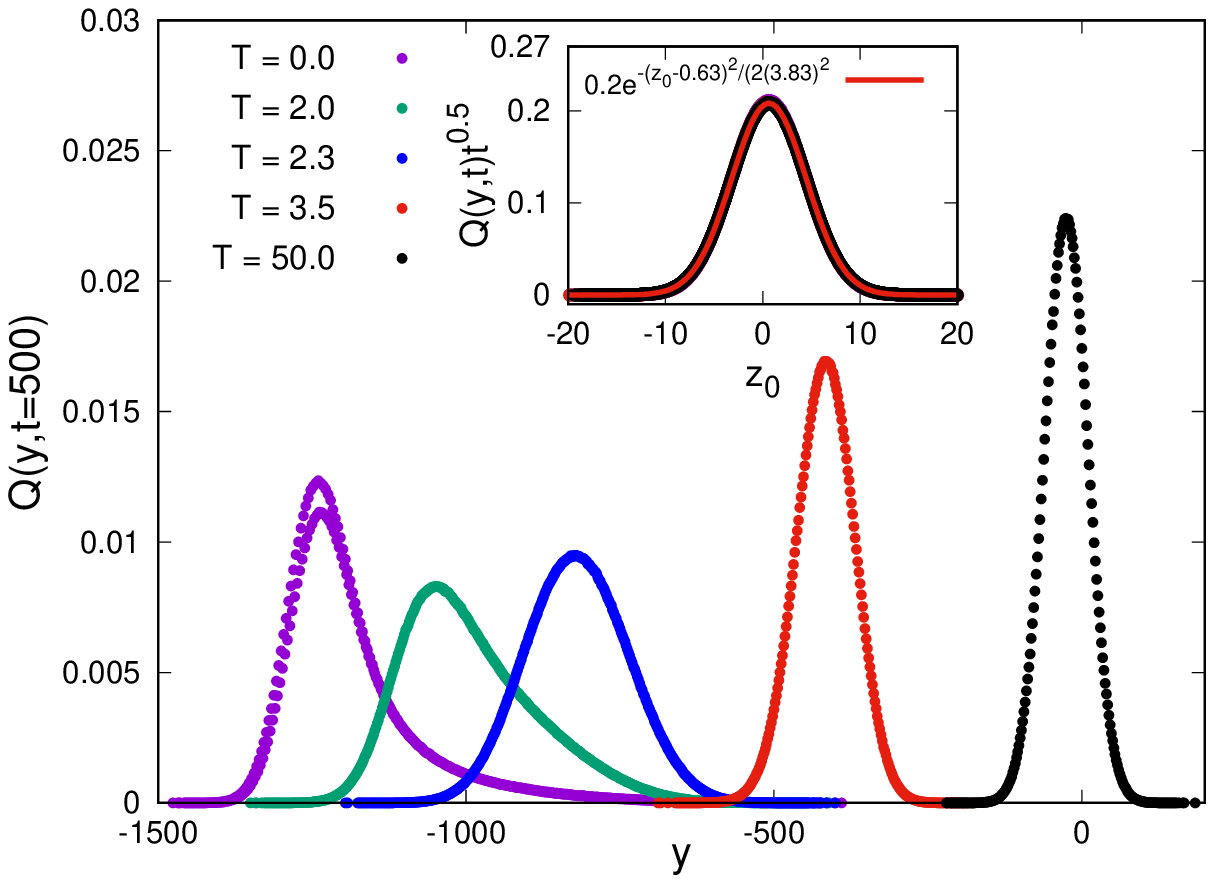} 
\caption{ Plot of distribution $Q(y,t)$ of displacement of the walker $y$ (where $y_i(t)= \sum_{\tau=0}^{t}E_i(\tau)$ and $E_i(t)$ is the nearest neighbor interaction energy for a particular site $i$ at time $t$) at different temperatures and at $t=500$. Inset shows data collapse of $Q(y,t)$ obtained using the scaling form given in Eq. (\ref{scaling of Q(y,t)}) at $3$ different times: $t=2000,4000,6000$ at $T=2.35$. The scaling function $g$ in Eq. (\ref{scaling of Q(y,t)}) is well described by a shifted Gaussian of the form, $g(z_0)= \exp(-\frac{(z_0-C)^2}{2\sigma_y^2})$, with the dimensionless scaled variable $z_0=\frac{y+B(T)t}{t^{\frac{1}{2}}}$. For $T=2.35$, $B(T)\approx  1.605$,$C\approx 0.631$ and $\sigma_y \approx 3.83$. }
\label{2D_Energy_walk}
\end{figure}

From the energy walk defined in Eq. (\ref{energy-walk}), we obtain an important quantity, namely the fluctuation of the energy walk displacement. By mapping the dynamics of nearest neighbor interaction energy for a particular site into a virtual energy-walk we obtained the distribution $Q(y,t)$ of the displacement $y$ of the walkers. At $T<T_c$, $Q(y,t)$ is observed to be a single peaked distribution peaked at $y=-y_m(T)$ and it is asymmetric about the most probable point $-y_m(T)$ [Fig. \ref{2D_Energy_walk}] and above $T_c$, $Q(y,t)$ is observed to be a biased Gaussian of the form $\sim t^{-1/2} \exp(-\frac{[y+B(T)t]^2}{t})$ (See inset of Fig. \ref{2D_Energy_walk}). Above $T_c$, $Q(y,t)$  follows a general scaling form
\begin{equation}
 Q(y,t) \sim t^{-\frac{1}{2}} g\left (\frac{y+B(T)t}{t^{\frac{1}{2}}}\right),
 \label{scaling of Q(y,t)}
\end{equation}
where $g(z_0)$ is the universal scaling function with $z_0=\frac{y+B(T)t}{t^{\frac{1}{2}}}$.

\vspace{2mm}
{\textit{Finite time scaling of diverging fluctuations for energy walk: }}

\vspace{3mm}

 We now focus on the fluctuation of the displacement of the energy walk. From the cumulative sum in Eq. (\ref{energy-walk}), the displacement of the walker associated with a given site at time $t$ is defined as 
\begin{equation}
\label{Definition of y}
    y_i(t) = \sum_{s=1}^{t} E_i(s),
\end{equation}
where $ E_i(s) =-S_i\sum_{j\in NN}S_j $, is the instantaneous local energy of $i$th site. The fluctuation of the displacement is therefore given by
\begin{equation}
    \sigma_y^2(t) = \langle y^2(t)\rangle - \langle y(t)\rangle^2.
\end{equation}
Using the definition of $y(t)$ in Eq. (\ref{Definition of y}), at large time in the scaling regime assuming stationarity, this can be written as 
\begin{equation}
\label{expression of fluctuation of y}
\sigma_y^2(t) \approx 2\sum_{\tau=1}^{t} (t-\tau)C_E(\tau),
\end{equation}
where $C_E(\tau) = \langle E(0)E(\tau)\rangle$ is the two time autocorrelation function of the local energy. The scaling form of $C_E(\tau)$ follows from the general dynamic scaling hypothesis for a local observable (see Eq. (\ref{energy autocorrelation}) of Appendix \ref{FTS of fluctuation}.2). Using the scaling dimension of the local energy density, $x_E = d-1/\nu$ (See Chapter 3, Sections 3.7 and 3.8 of Ref.\cite{Cardy}), for an Ising model of spatial dimension $d>1$, one can obtain a finite time scaling form for the energy walk fluctuation [Eq. (\ref{FTS of y}) of Appendix \ref{FTS of fluctuation}.2.] as
\begin{equation}
\label{finite time scaling of sigma_y^2}
    \sigma_y^2(t,T) \approx t^{[2-\frac{2}{z_c}(d-\frac{1}{\nu})]}\mathcal{H}\left[(T-T_c)t^{\frac{1}{\nu z_c}}\right],
\end{equation}
where the scaling function $\mathcal{H}$ arises from an integral over the rescaled time variable $u=\tau/t$, and it is explicitly given in the Appendix \ref{FTS of fluctuation}.2 [Eq. (\ref{scaling function H})].

 Above the critical temperature, the two time energy autocorrelation becomes exponential, which implies that the integral defining $\mathcal{H}$ is dominated by contributions from small values of the rescaled time variable $u$. As a result, as the critical point is approached from $T=T_c^+$, the energy walk fluctuation per unit time at large time [Eq. (\ref{divergence exponent for energy walk}) of Appendix \ref{FTS of fluctuation}.2] diverges as 
\begin{equation}
\label{divergence of energy walk fluctuation}
   \frac{\sigma_y^2}{t}\sim (T-T_c)^{-(\nu z_c-2d\nu+2)}.
\end{equation} 
Again,  using the known values of the exponents for $d=2$,  the divergence exponent governing the energy walk fluctuation becomes $\nu z-2d\nu +2 \approx 0.17$.

\begin{figure}[h]
\includegraphics[ width = \linewidth]{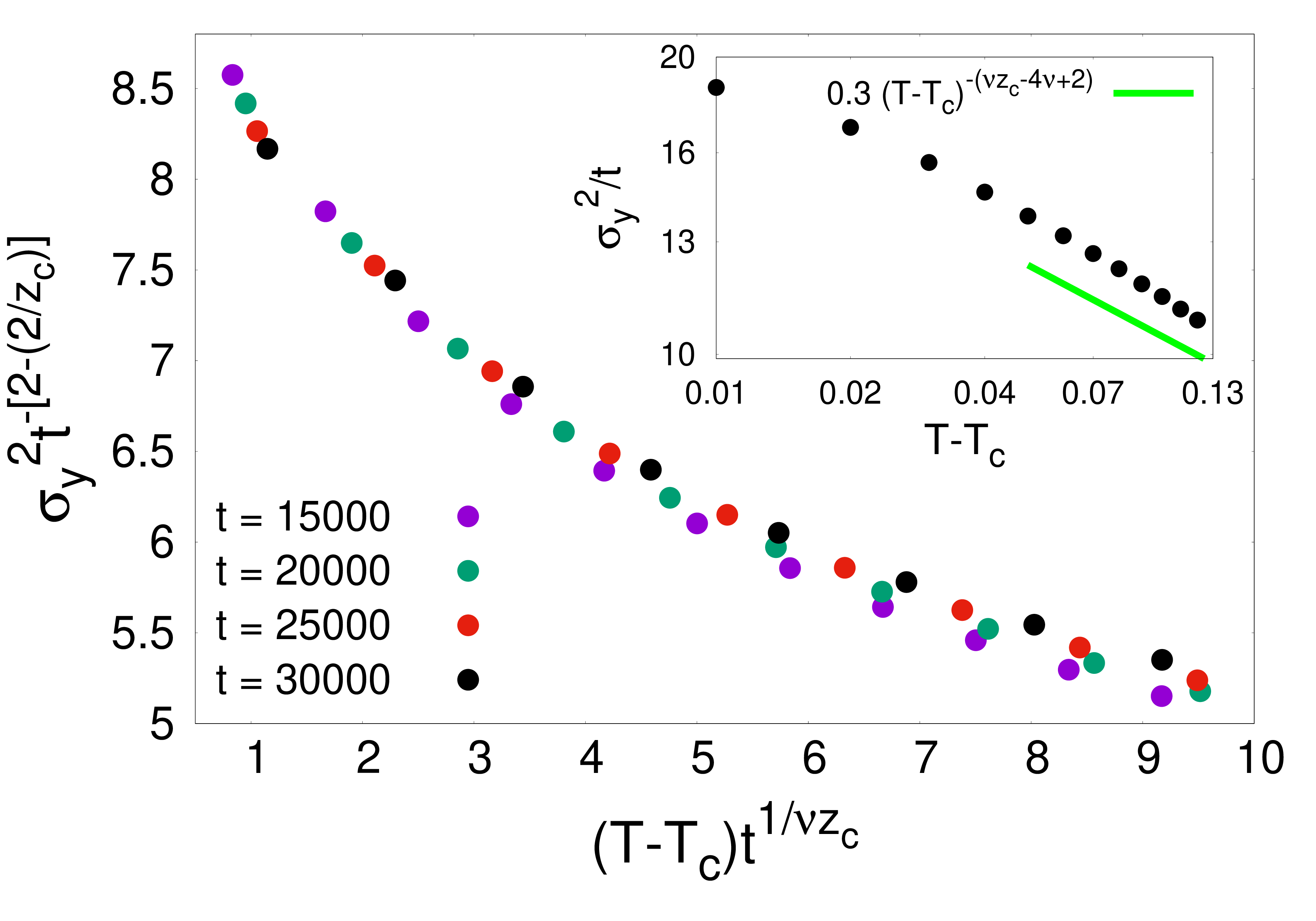} 
\caption{Data collapse of the scaled energy fluctuation $\sigma_y^2/t^{[2-(2/z_c)]}$ plotted against the scaled variable $(T-T_c)t^{1/\nu z_c}$ with $1/\nu z_c \approx 0.46$ at $T>T_c$. Results for four different evolution times $t=15000,20000,25000,30000$ collapse onto a single universal curve, confirming the validity of the finite time scaling relation proposed in Eq. (\ref{finite time scaling of sigma_y^2}). In inset, fluctuation of the energy walk displacement per unit time, $\sigma_y^2/t$, plotted as a function of $T-T_c$ on a  log-log scale at $t=30000$. As the critical point is approached from above, $\sigma_y^2/t$ diverges as $(T-T_c)^{-(\nu z_c-2d\nu+2)}$, and the slope extracted from the data is consistent with the expected exponent $\nu z_c-2d\nu+2 \approx 0.17$ [Eq. (\ref{divergence of energy walk fluctuation})]}. 
\label{fluctuation of y vs T}
\end{figure}

 The validity of finite time scaling relation given in Eq. (\ref{finite time scaling of sigma_y^2}) is confirmed numerically, as shown in Fig. \ref{fluctuation of y vs T}, where the simulation data collapse onto a single universal curve when plotted according to the proposed scaling form.

\section{Summary and Conclusions}

In summary, we have studied the behavior of the one and two dimensional Ising models following a  quench from a disordered configuration corresponding to a very high 
temperature to a lower temperature $T$ by tagging the spins. The dynamics of the tagged spins are regarded  as  walks performed in a virtual space. As expected, the nature of the walk changes as the critical temperature is crossed.  Another walk using the local energies is also studied in the two dimensional case.

In one dimension, the critical point is found following the analysis of the
ratio of $P(0,t)$ and the peak value $P(x_m,t)$ of the distribution as a function of time.  We also obtain 
 numerically the fluctuation of the Gaussian walks above $T_c=0$  that shows excellent agreement with the analytical estimate. 

 In two dimensions, we make more detailed studies. The critical temperature is estimated by the so called ratio method as well as
by calculating an appropriately defined Binder cumulant. Employing finite time scaling using the known values of the critical exponents, it is possible to  obtain excellent collapses of the scaled data at different times.

In this work, we therefore demonstrate that the complete set of equilibrium Ising critical exponents in two dimensions can be extracted within a finite time scaling framework using the walk representation, without invoking conventional finite size scaling. We claim that this method can be useful for
estimating the transition temperature and critical exponents in general. First, the transition temperature can be found from the crossing of the Binder cumulant curves at different times followed by a correct choice of $\nu z_c$ to obtain the collapse (Eq. \ref{FTS of binder cumulant}). The exponent $\beta$ can then be obtained from the collapse of the walk order parameter curves using Eq. \ref{finite time scaling of xm/t}. On the other hand, the finite time scaling of the energy fluctuations, given by Eq. \ref{finite time scaling of sigma_y^2} leads to an independent estimate of $z_c$ once $\nu z_c$ is known. Finally,  $\eta$ can be known from 
Eq. \ref{finite time scaling of sigma_x^2} and since more than two static critical exponents are obtained, one can check the scaling relations also. 

While the spin walk is closely connected to the studies made in \cite{Drouffe,Dornic,Godreche,Baldassarri}, the energy walk, so far not studied to the best of our knowledge, actually helps in making the study complete to provide a method of estimating  all the exponents, including static and dynamic.

\vspace{0.5 cm}
Acknowledgment: AP and PS acknowledge financial support by the Council of Scientific and Industrial Research, Government of India, through project no. $03/1495/23/$EMR-II.
SS acknowledges support from the Indian Academy of Sciences through the summer research fellowship program, under which he worked on this project in the Department of Physics,
University of Calcutta.  The authors also thank S. N. Majumdar for discussions.

\appendix

 \section{Derivation of the Finite time scaling of the Spin and Energy walk fluctuation}

\label{FTS of fluctuation}

In this Appendix, we present a detailed derivation of the finite-time scaling form of the fluctuation of both the spin and energy walk displacements in Ising models with spatial dimension $d\geq 1$, while the main text focuses on the cases $d=1$ and $d=2$. The derivation is based on the general dynamic scaling hypothesis for local observables and the scaling properties of the local spin and energy density. While only the final scaling form of $\sigma_y^2(T,t)$ and $\sigma_x^2(T,t)$ is presented in the main text, here we explicitly show how it follows from the scaling form of the spin and energy autocorrelation function and its integration over time.

 We start from the general dynamic scaling hypothesis for a local observable $A(r,t)$. Under a rescaling by a factor $b$, $r\to r/b,t\to t/b^{z_c},(T-T_c)\to (T-T_c)b^{1/\nu}$, and the observable transforms as $A\to b^{-x_A}A$, where $x_A$ is the scaling dimension of $A$. As a consequence, the correlation function satisfies \cite{Cardy} 
\begin{equation}
    C_A(r,t)=\langle A(r,t)A(0,0)\rangle = b^{-2x_A} F\left(\frac{r}{b},\frac{t}{b^{z_c}},(T-T_c)b^{\frac{1}{\nu}}\right).
\end{equation}
To obtain the temporal autocorrelation , we set $r=0$ and choose the rescaling factor $b$ such that $b^{z_c}=t$. This yields the scaling form
\begin{equation}
   C_A(t) = t^{-\frac{2x_A}{z_c}} F\left(\frac{t}{\xi_t}\right),
\end{equation}
where $\xi_t \sim(T-T_c)^{-\nu z_c}$, is the equilibration time.

\vspace{3mm}

\subsection{\textit{Spin walk}}

\vspace{3mm}

 From the cumulative sum in Eq. (\ref{walk-def}), the displacement of the spin walk associated with a given lattice site is defined as the sum of the local spin variable over time,
\begin{equation}
\label{x_sum}
    x(t) = \sum_{l=1}^{t} S(l).
\end{equation}
The fluctuation of the displacement is defined as 
\begin{equation}
    \sigma_x^2(t) = \langle x^2(t)\rangle - \langle x(t)\rangle^2.
\end{equation}
Using the definition of $x(t)$, we obtain
\begin{equation}
    \sigma_x^2(t) = \sum_{l=1}^{t}\sum_{l^{\prime}=1}^{t}\langle S(l)S(l^{\prime})\rangle -\langle S \rangle^2.
\end{equation}
Assuming stationarity, the spin spin correlation depends only on the time difference $\tau = l-l^{\prime}$. Defining the spin autocorrelation function as $C_S(\tau) = \langle S(0)S(\tau)\rangle - \langle S\rangle^2$, the fluctuation can be written as 
\begin{equation}
\label{spin fluctuation}
    \sigma_x^2(t) = 2\sum_{\tau=1}^{t-1}(t-\tau)C_S(\tau)+t,
\end{equation}
with $C_S(0) = 1$ by normalization.

\vspace{3mm}

\subsubsection{\textit{ Spin walk in the $d=1$ Ising model}}

\vspace{3mm}

 The exact two time spin correlation function for the one dimensional Ising model with Glauber dynamics is known from the solution of Glauber (see Eq. (76) of Ref. \cite{Glauber}) and is given by 
\begin{equation}
    \langle S_j(0)S_k(t)\rangle = \exp(-\alpha t) \sum_{l=-\infty}^{\infty}\eta^{|l+j-k|}I_{\ell}(\gamma \alpha t),
\end{equation}
where $I_{\ell}$ denotes the modified Bessel function, $\gamma = \tanh(2J/K_BT)$, $\eta = \tanh(J/K_B T)$ and $\alpha$ is the microscopic spin flip rate. For the single site autocorrelation ($j=k$), this reduces to
\begin{equation}
    C(t) = \langle S_j(0)S_j(t) \rangle = \exp(-\alpha t) \sum_{l=-\infty}^{\infty}\eta^{|l|}I_{\ell}(\gamma \alpha t).
\end{equation}
At large times, using the asymptotic form $I_{\ell}(x) \sim e^x/\sqrt{2 \pi x}$ for $x\to \infty$, one obtains 
\begin{equation}
    C(t) \sim \frac{\exp[-\alpha(1-\gamma)t]}{\sqrt{2\pi \gamma \alpha t}} \sum_{l=-\infty}^{\infty} \eta^{|l|}.
\end{equation}
Since, $|\eta|<1$ for all $T>0$, the sum converges to a finite constant, 
\begin{equation}
  \sum_{l=-\infty}^{\infty} \eta^{|l|} = 1+2\sum_{l=1}^{\infty} \eta^l = \frac{1+\eta} {1-\eta}.
\end{equation}
Thus the long time behavior of the autocorrelation function is 
\begin{equation}
\label{1D spin autocorrelation}
    C(t) \sim \frac{1}{\sqrt{t}} \exp\left(-\frac{t}{\xi_t}\right),
\end{equation}
with a finite inverse correlation time 
\begin{equation}
\label{1D correlation time}
    \xi_t^{-1} = \alpha(1-\gamma).
\end{equation}
This asymptotic form of the spin autocorrelation function is consistent with the large time decay obtained in Eq. (47) of Ref \cite{Brey}, where three distinct temporal regimes were identified.

In the limit of asymptotically large times $t\to \infty$, the discrete sum appearing in the definition of the walk displacement variance in Eq. (\ref{spin fluctuation})  can be replaced, to leading order, by a continuum integral using the Euler Maclaurin formula. Since we are interested in the dominant long time behavior, we retain only the exponential cutoff of the spin autocorrelation function given in Eq. (\ref{1D spin autocorrelation}), neglecting the subleading algebraic prefactor. This yields
\begin{equation}
    \sigma_x^2(t) \approx 2\int_{1}^{t}d\tau (t-\tau) \exp\left(-\frac{\tau}{\xi_t}\right).
\end{equation}
Introducing the scaled variable $\tau = tu$, the above expression takes the finite time scaling form
\begin{equation}
\label{FTS_1D_variance}
    \sigma_x^2(t) = t^{2}\mathcal{I}_1\left(\frac{t}{\xi_t}\right)=t^{2}\mathcal{I}_1\left(\alpha t(1-\tanh\left(\frac{2J}{K_BT}\right)\right),
\end{equation}
where the scaling function is given by
\begin{equation}
\label{1D scaling function}
    \mathcal{I}_1(y) = 2\int_{0}^{1}du(1-u)\exp(-yu).
\end{equation}
Since we are already in the asymptotic time regime (more precisely at $t>>\xi_t$), the exponential factor restricts the integral to $u\sim \xi_t/t <<1$. Introducing the rescaled variable 
$u=s\xi_t/t$ and approximating $1-u\approx 1$, the scaling function becomes
\begin{equation}
 \mathcal{I}_1\left(\frac{t}{\xi_t}\right) \approx 2\left(\frac{\xi_t}{t}\right)\int_{0}^{\infty}\exp(-s)ds.
\end{equation}
Evaluating the remaining integral yields 
\begin{equation}
  \mathcal{I}_1\left(\frac{t}{\xi_t}\right)\approx 2 \left(\frac{\xi_t}{t}\right).
\end{equation}
which finally leads to
\begin{equation}
\label{spin_walk_fluctuation_1D}
    \frac{\sigma_x^2(t)}{t} \sim \xi_t \sim \frac{1}{\left[1-\tanh\left(\frac{2J}{K_BT}\right)\right]}, \quad t>>\xi_t.
\end{equation}
Using the expansion $\tanh(x) \approx 1-2\exp(-2x)$ for large $x$, it follows that from Eq. (\ref{1D correlation time}) as $T\to 0^+$,
\begin{equation}
    \xi_t \sim \left(\frac{1}{2\alpha}\right) \exp\left(\frac{4J}{K_BT}\right),
\end{equation}
demonstrating an exponential divergence of the correlation time at zero temperature. Consequently, the spin walk displacement variance, which depends on $\xi_t$, also exhibits an exponential growth as $T\to0^+$.

\vspace{3mm}

\subsubsection{\textit{ Spin walk in the $d>1$ Ising model}}

\vspace{3mm}

 For the spin walk in an Ising model of spatial dimension $d>1$, the relevant observable is the local Ising spin operator, whose scaling dimension is $x_S = (d-2+\eta)/2$ (See Chapter 3, Sections 3.7 and 3.8 of Ref.\cite{Cardy}). Using the general result above, the spin autocorrelation function obeys
\begin{equation}
\label{spin autocorrelation}
    C_S(t) = \langle S(0)S(t)\rangle = t^{-\left[\frac{d-2+\eta}{z_c}\right]} F\left(\frac{t}{\xi_t}\right).
\end{equation}

For large observation times $t$ in the scaling regime, the second term in Eq. (\ref{spin fluctuation}) becomes irrelevant. retaining the leading contribution and using the scaling form of the spin autocorrelation given in Eq. (\ref{spin autocorrelation}), the discrete sum over $\tau$ can be converted to an integral in the large $t$ limit. Writing $\tau = tu$ we then obtain,
\begin{equation}
\label{sigma_x^2}
  \sigma_x^2(t) \approx t^{\left[2-\left(\frac{d-2+\eta}{z_c}\right)\right]} \left[ 2 \int_{0}^{1}(1-u)u^{-\left(\frac{d-2+\eta}{z_c}\right)}F\left(\frac{tu}{\xi_t}\right)du\right].
\end{equation}
This explicitly shows that the fluctuation of the energy walk displacement satisfies the finite time scaling form
\begin{equation}
\label{FTS of x}
\begin{split}
 \sigma_x^2(t,T) \approx t^{\left[2-\left(\frac{d-2+\eta}{z_c}\right)\right]} \mathcal{I}\left(\frac{t}{\xi_t}\right) \\ 
 \approx t^{\left[2-\left(\frac{d-2+\eta}{z_c}\right)\right]} \mathcal{I}\left[(T-T_c)t^{\frac{1}{\nu z_c}}\right],
 \end{split}
\end{equation}
where the scaling function $\mathcal{I}$ is defined by the following integral
\begin{equation}
\label{scaling function I}
  \mathcal{I}\left(\frac{t}{\xi_t}\right) = \left[ 2 \int_{0}^{1}(1-u)u^{-\left(\frac{d-2+\eta}{z_c}\right)}F\left(\frac{tu}{\xi_t}\right)du\right]. 
\end{equation}

 Above the critical temperature, the scaling function $F(t/\xi_t)$ becomes exponential i.e. $F(\tau/\xi_t)\sim \exp(-\tau/\xi_t) $ [See section 2.1 of Ref.\cite{Drouffe}], in this case the integral defining $\mathcal{I}$ is dominated by small values of $u$, since contributions from $\tau$ much larger than $\xi_t$ are exponentially suppressed. As a result, the scaling function $\mathcal{I}$ behaves as 
\begin{equation}
 \mathcal{I} \left(\frac{t}{\xi_t}\right)  \sim \left(\frac{\xi_t} {t}\right)^{\left[1-\left(\frac{d-2+\eta}{z_c}\right)\right]}.
\end{equation}
Substituting this behavior back into the finite time scaling form given in Eq. (\ref{FTS of x})  and using the divergence formula of the correlation time given we obtain
\begin{equation}
\label{divergence exponent for spin walk}
 \frac{\sigma_x^2}{t}\sim (T-T_c)^{-\left[\nu z_c-\nu(d-2+\eta)\right]}.   
\end{equation}
So, as the critical point is approached from $T=T_c^+$, the spin walk fluctuation per unit time at large observation time diverges with an exponent $\nu z_c-\nu(d-2+\eta)$.

\vspace{3mm}

\subsection{\textit{Energy walk}}

\vspace{3mm}

 Here we consider the energy walk in an Ising model of spatial dimension $d>1$. Since the relevant observable for the energy walk is the local energy density, whose scaling dimension is $x_E = d- \nu^{-1}$ (See Chapter 3, Sections 3.7 and 3.8 of Ref.\cite{Cardy}),being conjugate to the thermal scaling field. Using the general result above, the energy autocorrelation function obeys
\begin{equation}
\label{energy autocorrelation}
    C_E(t) = \langle E(0)E(t)\rangle = t^{-[\frac{2}{z_c}(d-\frac{1}{\nu})]} F\left(\frac{t}{\xi_t}\right).
\end{equation}
From the cumulative sum in Eq. (\ref{energy-walk}), the displacement of the walker associated with a given site at time $t$ is defined as 
\begin{equation}
\label{y_def}
    y_i(t) = \sum_{s=1}^{t} E_i(s).
\end{equation}
where $ E_i(s) =-S_i\sum_{j\in NN}S_j $, is the instantaneous local energy of $i$th site. The fluctuation of the displacement is therefore given by
\begin{equation}
\label{expression for energy fluctuation}
    \sigma_y^2(t) = \langle y^2(t)\rangle - \langle y(t)\rangle^2 = \sum_{s=1}^{t}\sum_{s^{\prime}=1}^{t}\langle E(s)E(s^{\prime})\rangle-\langle E\rangle^2.
\end{equation}
Introducing the time difference $\tau = s-s^{\prime}$, Eq. (\ref{expression for energy fluctuation}) can be rewritten as 
\begin{equation}
\label{energy_fluc}
\sigma_y^2(t) = 2\sum_{\tau=1}^{t} (t-\tau)C_E(\tau) + t,
\end{equation}
with $C_E(0) = 1$ by normalization.

 For large observation times $t$ in the scaling regime, the second term in Eq. (\ref{energy_fluc}) becomes irrelevant. retaining the leading contribution and using the scaling form of the energy autocorrelation given in Eq. (\ref{energy autocorrelation}), the discrete sum over $\tau$ can be converted to an integral in the large $t$ limit. Writing $\tau = tu$ we then obtain,
\begin{equation}
\label{sigma_y^2}
  \sigma_y^2(t) \approx t^{[2-\frac{2}{z_c}(d-\frac{1}{\nu})]} \left[ 2 \int_{0}^{1}(1-u)u^{-[\frac{2}{z_c}(d-\frac{1}{\nu})]}F\left(\frac{tu}{\xi_t}\right)du\right].
\end{equation}
This explicitly shows that the fluctuation of the energy walk displacement satisfies the finite time scaling form
\begin{equation}
\label{FTS of y}
\begin{split}
 \sigma_y^2(t,T) \approx t^{[2-\frac{2}{z_c}(d-\frac{1}{\nu})]} \mathcal{H}\left(\frac{t}{\xi_t}\right) \\ 
 \approx t^{[2-\frac{2}{z_c}(d-\frac{1}{\nu})]} \mathcal{H}\left[(T-T_c)t^{\frac{1}{\nu z_c}}\right],
 \end{split}
\end{equation}
where the scaling function $\mathcal{H}$ is defined by the following integral
\begin{equation}
\label{scaling function H}
    \mathcal{H}\left(\frac{t}{\xi_t}\right) = \left[ 2 \int_{0}^{1}(1-u)u^{-[\frac{2}{z_c}(d-\frac{1}{\nu})]}F\left(\frac{tu}{\xi_t}\right)du\right].
\end{equation}.

 Above the critical temperature, the scaling function $F(t/\xi_t)$ becomes exponential i.e. $F(\tau/\xi_t)\sim \exp(-\tau/\xi_t) $, in this case the integral defining $\mathcal{H}$ is dominated by small values of $u$, since contributions from $\tau$ much larger than $\xi_t$ are exponentially suppressed. As a result, the scaling function $\mathcal{H}$ behaves as 
\begin{equation}
 \mathcal{H} \left(\frac{t}{\xi_t}\right)  \sim \left(\frac{\xi_t} {t}\right)^{[1-\frac{2}{z_c}(d-\frac{1}{\nu})]}.
\end{equation}
Substituting this behavior back into the finite time scaling form given in Eq. (\ref{FTS of y})  and using the divergence formula of the correlation time given we obtain
\begin{equation}
\label{divergence exponent for energy walk}
 \frac{\sigma_y^2}{t}\sim (T-T_c)^{-(\nu z_c-2d\nu+2)}.  
\end{equation}
So, as the critical point is approached from $T=T_c^+$, the energy walk fluctuation per unit time at large observation time diverges with an exponent $\nu z_c-2d\nu+2$.

\section{Finite time scaling of the Binder cumulant}
\label{app:B}

In this appendix, we briefly derive the finite time scaling form of the Binder cumulant of the spin walk displacement.

we  perform a coarse graining transformation in which spatial lengths are rescaled as $L\to L^{\prime} = L/b$ with a scale factor $b>1$. Since, time and lengths are related dynamically, this induces a rescaling of time $t\to t^{\prime} = t/b^{z_c}$. The deviation from criticality $\epsilon = T-T_c$ rescales as $\epsilon \to \epsilon^{\prime} = \epsilon b^{1/\nu}$, where $\nu$ is the correlation length exponent.

The Binder cumulant $U(T,t)$, defined in Eq. (\ref{binder}), is a dimensionless quantity constructed from ratios of moments of the spin walk displacement. Consequently, all anomalous scaling dimensions cancel, and $U(T,t)$ is invariant under the above scale transformation. Assuming homogeneity, it therefore satisfies
\begin{equation}
    U(T,t) = U\left(\frac{t}{b^{z_c}},\epsilon b^{\frac{1}{\nu}}\right)
\end{equation}
Choosing the scale factor $b=t^{1/z_c}$ yields the finite time scaling form
\begin{equation}
\label{FTS_binder}
    U(T,t)  = \mathcal{F}\left[(T-T_c)t^{\frac{1}{\nu z_c}}\right] = \mathcal{F}\left(\frac{t}{\xi_t}\right),
\end{equation} 
where $\xi_t\sim |T-T_c|^{\nu z_c}$, is the equilibration time.

\vspace{6mm}

\section{Finite time scaling of the spin walk order parameter }

\label{scaling of order parameter}

In this appendix, we derive the finite time scaling form of the spin walk order parameter 
$x_m(t)/t$ in the vicinity of the critical temperature $T_c$.

Near criticality, the system is characterized by a growing dynamical length scale $L(t)$, which represents the largest distance over which spins have dynamically correlated up to time $t$. Close to $T_c$ and for $T<T_c$, the dynamic length scale grows as
\begin{equation}
 L(t)\sim t^{1/z_c} \quad \text{for}\quad  t<<t_{eq}, 
\end{equation}
after which it saturates at the equilibrium correlation length $\xi\sim |T-T_c|^{-\nu}$.

 The coarse graining transformation that rescales space, time, and the deviation from criticality has already been introduced in Appendix A. Using the same transformation rules, we now focus on the scaling behavior of the spin walk order parameter.

Close to $T_c$, we assume that the spin walk order parameter $x_m(t)/t$ behaves as a homogeneous function under the above scaling transformation. Thus under coarse graining 
\begin{equation}
    \frac{x_m}{t}(t,\epsilon) = b^q\frac{x_m}{t}\left(\frac{t}{b^{z_c}},\epsilon b^{\frac{1}{\nu}}\right),
\end{equation}
where $q$ is an a priori unknown scaling exponent associated with $x_m/t$.

We now choose the scale factor $b$ such that the rescaled time is of order unity, $b=t^{1/z_c}$. Substituting this choice into the homogeneity relation yields the finite time  scaling form 
\begin{equation}
  \frac{x_m}{t}(t,\epsilon) = t^{\frac{a}{z_c}} \mathcal{G}\left(\epsilon t^{\frac{1}{\nu z_c}}\right),
\end{equation}
where $\mathcal{G}$ is a universal scaling function.

In the long time limit $t\to \infty$, for temperatures below criticality, the spin walk order parameter $x_m/t$ vanish exactly in the same way as equilibrium magnetization [Eq. (\ref{walk order parameter})]. This requirement implies that for large argument 
\begin{equation}
    \mathcal{G}(y)\sim y^{\beta} \quad \text{as} \quad y\to \infty.
\end{equation}
Consistency with the finite time scaling form then fixes the exponent to be $a=-\beta/\nu$.

We thus arrive at the final finite time scaling relation for the spin walk order parameter,
\begin{equation}
\frac{x_m}{t}(t,T) = t^{-\frac{\beta}{\nu z_c}} \mathcal{G}\left[(T_c-T)t^{\frac{1}{\nu z_c}}\right].
\label{FTS of x_m/t}
\end{equation}
At the critical point $T=T_c$, this implies a pure power law decay
\begin{equation}
    \frac{x_m}{t}(t) \sim t^{-\frac{\beta}{\nu z_c}},
\end{equation}
while away from criticality the crossover is governed by the equilibration time 
\begin{equation}
    t_{eq} \sim (T_c-T)^{-\nu z_c}.
\end{equation}

\end{document}